\documentclass[acmsmall,screen]{acmart}
\AtBeginDocument{%
  }

\setcopyright{acmlicensed}
\copyrightyear{2018}
\acmYear{2018}
\acmDOI{XXXXXXX.XXXXXXX}
\acmConference[Conference acronym 'XX]{Make sure to enter the correct
  conference title from your rights confirmation email}{June 03--05,
  2018}{Woodstock, NY}
\acmISBN{978-1-4503-XXXX-X/2018/06}

\usepackage{algorithm2e}
\usepackage{enumitem}
\usepackage{array}
\usepackage{wrapfig}
\usepackage{amsmath,amsfonts}
\usepackage[noend]{algpseudocode}
\usepackage{graphicx}
\usepackage{textcomp}
\usepackage{float}
\usepackage{listings}
\usepackage{xspace}
\usepackage{multirow}
\usepackage{amsthm}

\usepackage{balance}
\usepackage{algpseudocode}
\usepackage{colortbl}

\usepackage[skins]{tcolorbox}
\usepackage{xcolor,pifont}
\usepackage{multicol}
\newcommand*\colourcheck[1]{%
	\expandafter\newcommand\csname #1check\endcsname{\textcolor{#1}{\ding{52}}}%
}
\colourcheck{blue}
\colourcheck{green}
\colourcheck{red}
\newtcolorbox{boxB}[2][]{%
  enhanced,colback=white,colframe=black,coltitle=black,
  sharp corners,
  toprule=1.0pt,
  rightrule=0.3pt,
  leftrule=0pt,
  bottomrule=0pt,
  fonttitle=\itshape\scshape\large,
  left=0pt,right=5pt,top=5pt,bottom=3pt,
  attach boxed title to top right={yshift=-0.3\baselineskip-0.4pt,xshift=-5mm},
  boxed title style={tile,size=minimal,left=0.2mm,right=0.5mm,
    colback=white,before upper=\strut},
  title=#2,#1
}

\newcommand{\tool}{\textsc{Cerberus}\xspace}
\newcommand{\model}{\textsc{PE}\xspace}

\newboolean{showcomments}
\setboolean{showcomments}{true}
\ifthenelse{\boolean{showcomments}}
 { \newcommand{\mynote}[2]{
      \fbox{\bfseries\sffamily\scriptsize#1}
        {\small$\blacktriangleright$\textsf{\emph{#2}}$\blacktriangleleft$}}}
        { \newcommand{\mynote}[2]{}}

\usepackage{tikz}

\newcolumntype{L}[1]{>{\raggedright\arraybackslash}p{#1}}

\newcommand{\code}[1]{{\footnotesize\texttt{#1}}}
\usepackage{amsthm}
\definecolor{dkgreen}{rgb}{0,0.6,0}
\definecolor{gray}{rgb}{0.5,0.5,0.5}
\definecolor{lightgray}{rgb}{211, 211, 211}
\definecolor{mauve}{rgb}{0.58,0,0.82}

\definecolor{dkgreen}{rgb}{0,0.6,0}
\definecolor{custom-red}{rgb}{1,0,0}
\definecolor{custom-blue}{rgb}{0,0,1}
\definecolor{gray}{rgb}{0.5,0.5,0.5}
\definecolor{mauve}{rgb}{0.58,0,0.82}

\definecolor{c1}{HTML}{f4cccc}
\definecolor{c2}{HTML}{f5cdcd}
\definecolor{c3}{HTML}{fffcfc}
\definecolor{c4}{HTML}{ffffff}
\definecolor{c5}{HTML}{ffffff}
\definecolor{c6}{HTML}{fffdfd}
\definecolor{c7}{HTML}{f5cfcf}
\definecolor{c8}{HTML}{fffbfb}
\definecolor{c9}{HTML}{ffffff}
\definecolor{c10}{HTML}{fffdfd}
\definecolor{c11}{HTML}{fefafa}
\definecolor{c12}{HTML}{fef7f7}
\definecolor{c13}{HTML}{ffffff}
\definecolor{c14}{HTML}{fffefe}
\definecolor{c15}{HTML}{ffffff}
\definecolor{c16}{HTML}{fefafa}
\definecolor{c17}{HTML}{fdf3f3}
\definecolor{c18}{HTML}{fffefe}
\definecolor{c19}{HTML}{fdf5f5}
\definecolor{c20}{HTML}{ffffff}

\begin{document}


\title[{\tool}: Multi-Agent Reasoning and Coverage-Guided Exploration for Static Detection of Runtime Errors]{{\tool}: Multi-Agent Reasoning and Coverage-Guided Exploration for Static Detection of Runtime Errors}




\setcopyright{acmcopyright}
\copyrightyear{2018}
\acmYear{2018}
\acmDOI{X.Y}

\settopmatter{printacmref=false, printfolios=false}


\author{Hridya Dhulipala}
\orcid{0009-0001-4474-2984}
\affiliation{%
  \institution{University of Texas at Dallas}
  \city{Dallas}
  \country{USA}
}
\email{hridya.dhulipala@utdallas.edu}

\author{Xiaokai Rong}
\orcid{0009-0000-8457-8528}
\affiliation{%
  \institution{University of Texas at Dallas}
  \city{Dallas}
  \country{USA}
}
\email{xiaokai.rong@utdallas.edu}

\author{Tien N. Nguyen}
\orcid{0009-0006-7962-6090}
\affiliation{%
  \institution{University of Texas at Dallas}
  \city{Dallas}
  \country{USA}
}
\email{Tien.N.Nguyen@utdallas.edu}









\begin{abstract}
In several software development scenarios, it is desirable to detect runtime
errors and exceptions in code snippets {\em without actual execution}. A typical example is to detect runtime exceptions in online code snippets before integrating them into a codebase.
In this paper, we propose {\tool}, a novel predictive, execution-free coverage-guided testing
framework. 
{\tool} uses LLMs to generate the inputs that trigger runtime errors and to perform code
coverage prediction and error detection without code
execution. With a two-phase feedback loop, {\tool} first aims to both increasing code coverage and detecting runtime errors, then shifts to focus only detecting runtime errors when the coverage reaches 100\% or its maximum, enabling it to perform better than prompting the LLMs for both purposes.
%
Our empirical evaluation demonstrates that
{\tool} performs better than conventional and learning-based testing frameworks for (in)complete code snippets by generating high-coverage test cases more efficiently, leading to the discovery of more runtime errors.
\end{abstract}

\keywords{Runtime Error Detection, Multi-Agent Reasoning, Coverage Exploration}



\maketitle

\section{Introduction}
\label{sec:intro}

Detecting runtime errors in code snippets without execution is crucial in software development. Online forums like StackOverflow (S/O) serve as valuable resources but often contain code with defects and runtime errors that can introduce risks to projects adopting them. Hong {\em et al.}~\cite{hong21dicos} reviewed +1.9 million S/O posts and identified 14,719 insecure code snippets that had been adopted into 151 popular open-source projects. Therefore, tools that help in the early detection of runtime errors in code snippets from online forums are essential, as they mitigate the costs associated with fixing faulty code later in the software cycle. Beyond this, such tools can also function as programming assistants, helping address runtime errors early, even before testing and execution. Thus, such tools reduce the risk of undetected bugs affecting deployed software.

To address these issues, dynamic analysis and dynamic symbolic execution approaches ~\cite{GLS:PLDI05,SMA:FSE05,CDE:OSDI08} provide more accurate results by analyzing actual runtime behavior, but they require executable code and thus are inapplicable to incomplete  snippets. Static analysis and symbolic execution approaches ~\cite{Cole-2006,10.1016/S0167-6423(01)00013-2} have been used as alternatives to dynamic testing, since these techniques examine program semantics without executing the code, making them suitable for analyzing non-executable snippets. However, they suffer from high false positive rates due to overestimation.

A coverage-guided testing framework~\cite{JQF,10.1145/3133956.3138820} systematically discovers software defects by executing test inputs on a target program. The process begins with {\em seed corpus generation}, where an initial set of test cases is created from existing test suites, real-world inputs, or example code snippets. 
The test cases are {\em executed} on the program under test while the framework monitors code coverage, tracking~which parts of the code have been covered. If an input triggers unexpected behavior, it is logged as a {\em potential defect} for further analysis. This process iteratively refines test inputs through coverage feedbacks, prioritizing those that explore untested paths or expose~errors.

Advances in machine learning (ML)~\cite{vuldeepecker18,wu2022vulcnn,yioopsla19} have introduced alternative methods for defect detection by identifying buggy code patterns and triggering those bugs by relevant test cases. 
Large language models (LLMs)  have also been used to generate test cases to detect runtime errors based on their ability to reason about code semantics~\cite{fuzz4all}. However, our preliminary experiments (Section~\ref{sec:vanilla-prompt}) indicate that naively prompted LLMs struggle to accurately identify runtime errors, achieving only 31\% precision and 33\% recall for Java. This limitation arises from their lack of execution knowledge, as LLMs do not run code, but rather infer potential defects based on learned~patterns~\cite{forge24}.

\begin{figure*}[t]
\begin{center}
\includegraphics[width=5.2in]{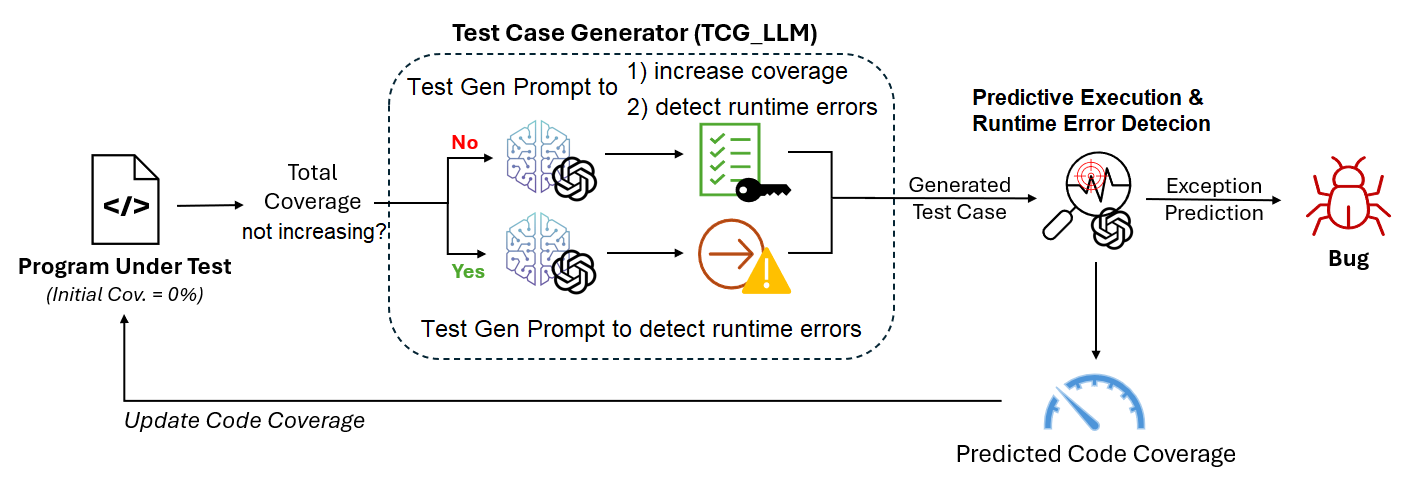}
\vspace{-12pt}  
\caption{Predictive Coverage-Guided Testing with Feedback Loop and Condition-based Dual Prompting}
\label{fig:fuzzwise}
\end{center}
\end{figure*}

All the above approaches (traditional frameworks, machine learning and LLM-based test case generation) share a fundamental limitation: they require the test cases to be executed on the code. If the code is non-executable, they fail to detect runtime errors. Thus, we propose {\tool}, {\em an LLM-based {\bf multi-agent framework with a two-phase iterative feedback loop}} (Fig.~\ref{fig:fuzzwise}) that can be used to effectively detect unhandled runtime errors in (in)complete code. Since directly prompting an LLM for runtime error detection yielded suboptimal results (Section~\ref{sec:vanilla-prompt}), {\tool} is designed with the principle of coverage-guided testing framework. 
We improve the guidance provided to the LLM for exception detection through a structured two-phase prompting mechanism.






{\tool}' framework consists of two main components: test case generation and predictive execution. However, standard direct prompts for both components in generating test cases often resulted in low-diversity inputs that failed to sufficiently explore new execution paths, limiting their usefulness for bug detection (Section~\ref{sec:rq4}). To this end, {\tool} employs a novel {\bf condition-based dual-prompting} strategy. This method dynamically switches between two distinct strategies: one specifically aimed both {\em increasing code coverage} and {\em maximizing the likelihood of triggering runtime errors}, and another aimed to {\em focusing only on error detection}. The decision to switch prompts is guided by the cumulative code coverage achieved in the previous iteration—if coverage remains increasing, the LLM is instructed to generate additional test cases to {\em both expand the coverage and detect errors}. If a generated test case improves coverage and exposes a potential error, {\tool} refines test cases for further error detection. If code coverage is predicted to not increase or reaches 100\%, {\tool} prompts the LLM to shift its focus only on runtime error detection. This iterative feedback allows the LLM to operate in a more structured manner, improving effectiveness over prompting with two focuses at the same time.


In the second component, we leverage LLMs to perform the prediction of code execution for the generated test case with two main tasks: 1) runtime error prediction, and 2) code coverage prediction. This component plays the role of code execution in the traditional process in executing the program with a test case and measure the code coverage.
This allows the system to simulate test execution without actually running the code, bridging the gap between static analysis and {\tool}. For the second task, we leverage CodePilot~\cite{forge24} with Chain-of-Thought prompting.

Our experiments on the FixEval dataset~\cite{haque2023fixeval} indicate that, within a specified time limit, {\tool} generates significantly fewer test cases while detecting more runtime errors compared to the conventional fuzzer Jazzer~\cite{jazzer} and LLM-based Fuzz4All~\cite{xia2024fuzz4all}. 
Our results show that it achieves rapid coverage improvement with noticeably shorter periods of coverage plateau compared to the baselines. In the case of incomplete code, while all existing approaches fail, {\tool} is able to actively generate test cases and detect exceptions. With a precision of 58\%, nearly two-thirds of the detected runtime errors are predicted accurately. The recall of 41\% suggests that it can identify 4 out of 10 possible runtime errors in the ground truth. Furthermore, we demonstrate its effectiveness in detecting runtime errors in actual SO code snippets. {\tool} also supports for both Java and Python. In brief, the key contributions of this paper include:

{\bf 1. {\tool}:} two-phase, coverage-guided testing for (in)complete code without execution. 


{\bf 2. A two-phase, multi-agent framework for LLM Test Generation and Coverage Prediction:} for detecting runtime errors.

{\bf 3. Empirical Evaluation} showing {\tool} is more effective and efficient than the baselines.


\section{Motivation: The Challenges of Runtime Error Detection}
\label{sec:vanilla-prompt}


\begin{wrapfigure}{r}{0.49\textwidth}
\begin{minipage}{0.48\textwidth}
	\centering
	\lstset{
		numbers=left,
		numberstyle= \tiny,
		keywordstyle= \color{blue!70},
		commentstyle= \color{red!50!green!50!blue!50},
		frame=shadowbox,
		rulesepcolor= \color{red!20!green!20!blue!20} ,
		xleftmargin=1.5em,xrightmargin=0em, aboveskip=1em,
		framexleftmargin=1.5em,
                numbersep= 5pt,
		language=Java,
    basicstyle=\scriptsize\ttfamily,
    numberstyle=\scriptsize\ttfamily,
    emphstyle=\bfseries,
                moredelim=**[is][\color{red}]{@}{@},
		escapeinside= {(*@}{@*)}
	}
\begin{lstlisting}[]
(*@{\color{orange}{VANILLA PROMPT} To predict Runtime Errors.@*)
Given the following JAVA program, List all the unique Runtime Exceptions that can be triggered in the program. Use the following format ...
Possible Runtime Exceptions - 
<Runtime Exception 1>
<Runtime Exception 2> 
\end{lstlisting}
\end{minipage}
\vspace{-32pt}
\begin{minipage}{0.48\textwidth}
	\centering
	\lstset{
		numbers=left,
		numberstyle= \tiny,
		keywordstyle= \color{blue!70},
		commentstyle= \color{red!50!green!50!blue!50},
		frame=shadowbox,
		rulesepcolor= \color{red!20!green!20!blue!20} ,
		xleftmargin=1.5em,xrightmargin=0em, aboveskip=1em,
		framexleftmargin=1.5em,
                numbersep= 5pt,
		language=Java,
    basicstyle=\scriptsize\ttfamily,
    numberstyle=\scriptsize\ttfamily,
    emphstyle=\bfseries,
                moredelim=**[is][\color{red}]{@}{@},
		escapeinside= {(*@}{@*)}
	}
\begin{lstlisting}[]
import java.util.*;
public class Main {
  public static void main(String[] args){
    Scanner sc = new Scanner(System.in);
    int n = sc.nextInt();
    int k = sc.nextInt();
    int as[] = new int[n];
    for (int i = 0; i < n; i++) as[i] = sc.nextInt();
    for (int i = 0; i < k && (k+i) < n; i++) {
      if (as[k+i] > as[i]) System.out.println(x:''Yes'');
      else System.out.println(x:''No'');
    }
  }
}  
\end{lstlisting}
\vspace{-15pt}
\caption{An Example of a Target Java Code}
\label{fig:motiv}
\end{minipage}
\end{wrapfigure}


An intuitive approach to statically detect runtime errors is to leverage LLMs via direct prompting, but our experiment shows this to be unreliable due to frequent inaccuracies.

\subsubsection*{\bf A Case Study in LLM Misdiagnosis} 
Consider the example in Fig.~\ref{fig:motiv}. When prompted with a request: {\em"Find as many runtime errors as possible in the given code ..."}, GPT-4 incorrectly detects  \code{ArrayIndexOutOfBoundsException} with the explanation:

"{\em (1) The condition in the \code{for} loop, \code{for (int i = 0; i < k $\&\&$ (k + i) < n; i++)}, tries to ensure that the indices accessed do not go out of bounds. However, condition \code{k+i < n} does not prevent \code{i} from going out of bounds for the array \code{as[ ]} at index \code{as[i]}.

(2) If \code{k} is larger than \code{n}, \code{i} will access an invalid index in \code{as[ ]} during the first few iterations (i.e. \code{as[i]} might access an index outside the range of the array \code{as[ ]}), resulting in \code{ArrayIndexOutOfBoundsException}.

Example: If \code{n=5} and \code{k=6}, the first iteration (\code{i=0}) will attempt to access \code{as[0]} and \code{as[6]}, which results in an out-of-bounds error because the valid array indices are 0 to 4.}"

Both GPT's reasoning and its example are {\bf incorrect}. For (1), the condition \code{(k+i) < n} prevents an index out-of-bound error in \code{as[i]}. Since \code{k + i < n}, it follows that  \code{i < n - k}. Given \code{k > i} (first loop condition) and \code{i >= 0}, we conclude that 0 <= \code{i < n}, ensuring as \code{as[i]} never causes \code{ArrayIndexOutOfBoundsException}. In GPT's example (\code{n=5} and \code{k=6}), the second loop condition \code{k}+\code{i} < \code{n} is false in the first iteration (i.e., \code{6}+\code{0} < \code{5} is false), preventing execution inside the loop and avoiding an exception. Reasoning (2) is also incorrect: if \code{k}>\code{n}, then \code{k}+\code{i} > \code{n}, violating the second-loop condition, ensuring that \code{as[k + i]} is never accessed. Thus, \code{ArrayIndexOutOfBoundsException} would not occur. This shows how LLMs with simple prompts misinterprets code logic, leading to false positives and inaccurate error predictions.

\subsubsection*{\bf Preliminary Experiment}

To quantify the effectiveness of LLM-based error detection, we conducted an experiment using GPT-4o ~\cite{GPT} with vanilla prompting (Fig.~3) and an extended approach in which it was also asked to generate test inputs that trigger predicted errors. The results (Tables~\ref{tab:pre1}--\ref{tab:pre2}) indicate the following:

(1) Vanilla prompting achieves a precision of 0.32, a recall of 0.312, and an F1-score of 0.316.

(2) When extended to include test case generation, performance declines further, with a precision of 30\% and an F1-score of 30.3\%.

This decline is due to the following two main issues:

(1) {\em False positives} – The LLM frequently misidentifies exceptions, such as \code{IOException} (incorrectly predicted to be triggered in at least 34 of the 100 data samples) and \code{NumberFormatException} (incorrectly predicted to be triggered in at least 42 of the 100 data samples), likely due to overgeneralization from learned patterns.

(2) {\em False negatives} - Although an error is identified at runtime, model-generated test cases often fail to trigger that error.

In contrast, as seen in Tables~\ref{tab:pre1} and \ref{tab:pre2}, our approach, {\tool}, achieves higher accuracy (see Section~\ref{sec:eval} for details).




\begin{table}[t]
  \centering
\begin{minipage}{0.49\textwidth}
\captionsetup{type=table}
 \captionof{table}{Vanilla Prompt on complete code} 
  \label{tab:pre1}  
  \vspace{-5pt}
    \tabcolsep 3.2pt
   \scalebox{0.9}{  
      \begin{tabular}{l|c|c|c}  
        \toprule  
        \multirow{2}{*}{\textbf{Vanilla Prompting}} & \multicolumn{3}{c}{\textbf{Evaluation Metrics}} \\ \cline{2-4}  
        & \textit{Prec} & \textit{Rec} & \textit{F1-Score} \\ \hline  
        RT Error Det. & 0.32 & 0.312 & 0.316 \\  
        RT Error Det. + Trig. Input & 0.30 & 0.306 & 0.303 \\ \hline  
        \textbf{\tool} & \textbf{0.72} & \textbf{0.56} & \textbf{0.63} \\  
        \bottomrule  
      \end{tabular}  
    } 
\end{minipage}%
\hfill
\begin{minipage}{0.49\textwidth}
\captionsetup{type=table}
 \captionof{table}{Vanilla Prompt on Incomplete Code}
\label{tab:pre2} 
\vspace{-5pt}
\tabcolsep 3.2pt
\scalebox{0.9}{  
      \begin{tabular}{l|c|c|c}  
        \toprule  
        \multirow{2}{*}{\textbf{Vanilla Prompting}} & \multicolumn{3}{c}{\textbf{Evaluation Metrics}} \\ \cline{2-4}  
        & \textit{Prec} & \textit{Rec} & \textit{F1-Score} \\ \hline  
        RT Error Det. & 0.315 & 0.3 & 0.307 \\  
        RT Error Det. + Trig. Input & 0.29 & 0.289 & 0.289 \\ \hline  
        \textbf{\tool} & \textbf{0.58} & \textbf{0.41} & \textbf{0.48} \\  
        \bottomrule  
      \end{tabular}  
    }
\end{minipage}
\end{table}

\subsubsection*{\bf Traditional Fuzzing: A Partial Solution with Execution Constraints}
Before LLMs, traditional coverage-guided fuzz testing has long been employed for error detection. However, this approach comes with notable inefficiencies:

1. {\em Excessive test case generation} – Even when execution is constrained to a 5-minute window, our evaluation found that over {\bf 7 million} test cases were produced, yet {\bf less than half} contributed to new code coverage.
 
2. {\em Execution dependency} – Traditional fuzzers require a fully executable codebase, meaning that they are ineffective for incomplete or non-compilable snippets, such as those frequently found in early development stages or extracted from Stack Overflow.

\subsubsection*{\bf ML-Based Fuzzing: A Step Forward, but Still Execution Dependent}
To address traditional fuzzing inefficiencies, ML-based fuzzers such as CODAMOSA~\cite{10.1109/ICSE48619.2023.00085} Fuzz4All ~\cite{xia2024fuzz4all} (included as a baseline in our study) aim to generate smarter, targeted test cases. However, despite their advancements, these approaches still require the code to be executable, limiting their applicability. 

\subsubsection*{\bf Few-shot Prompting and Chain-of-Thought (CoT)} In few-shot learning, we prompted GPT-4o with test case exemplars for specific exceptions (e.g., \code{NullPointerException}). However, this led to overfitting on exemplars—LLMs often mimicked the input patterns, reused hardcoded values or similar API calls, failing to adapt to code-specific semantics or effectively explore the input space.

We also used CoT for the GPT-4o. In our experiment, CoT prompts often produced verbose rationales instead of concise, executable test cases. Additionally, the reasoning chains increased token usage without improving exception detection rates.

\section{Key Ideas and Approach Overview}

We design {\tool}
with the following key ideas:


\subsubsection*{An {\bf Multi-Agent} framework for Execution-free Coverage-guided Test Generation}



{\tool} employs a multi-agent framework of LLMs, each LLM serving distinct roles: (1) a Test Case Generator (TCG) that produces diverse test inputs and (2) Predictive Executor (PE), responsible for code coverage prediction and runtime error detection. The TCG collaborates with PE to iteratively refine test cases, improving their effectiveness. While TCG aims to maximize test space exploration to enhance recall, this can lead to reduced precision. To mitigate this, PE assesses the quality of generated test cases, ensuring they target uncovered statements and detect runtime errors. Redundant test cases are discarded to maintain efficiency.



\subsubsection*{Integration of {\bf Two-phase, Feedback Loop} and {\bf Condition-based
Dual-prompting Strategy}}




Our experiments (see Baselines in Section~\ref{sec:rq1}) demonstrated that a naive simulation of coverage-guided testing framework with LLMs yielded suboptimal results due to various limitations (discussed in Section~\ref{sec:rq1}). To address this, we implemented a condition-based dual-prompting strategy, where each prompt in the test generation process serves a distinct goal. 
In the first phase, the LLM is prompted to generate test cases to target both goals 1) increasing coverage, and 2) detecting runtime errors (the ``No'' label in Fig.~\ref{fig:fuzzwise}).
Once the total coverage does not increase for certain amount of time or already reaches 100\%, 
the second phase prioritizes test cases aimed at triggering runtime errors (the ``Yes'' label in Fig.~\ref{fig:fuzzwise}). {\em This structured prompting approach ensures the LLM for TCG optimizes and focuses on test case generation for runtime error detection while the total coverage might not increase}. In fact, when 
{\em merging them into a single phase (one prompt with two goals)}, we found that this confuses the LLMs, making it {\em less effective} in runtime error detection (Section~\ref{sec:rq1}).


\subsubsection*{\bf Dual Tasks of Code Coverage Prediction and Runtime Error Detection}


We leverage LLM to build the predictive execution (PE) component, which performs the prediction of code execution with the generated test case. The PE component plays the role of {\em both runtime error prediction and code coverage prediction}. 
For the second task of code coverage prediction, we use the Chain-of-Thought (CoT) prompting from CodePilot~\cite{forge24}, a coverage prediction approach. CodePilot utilizes a prompting strategy that guides the LLM to develop a step by step reasoning based plan to predict the execution steps and resulting code coverage of a test case.
The PE component also instructs the LLM to analyze whether the generated test case is likely to trigger an exception. By assuming dual responsibilities—code coverage prediction and error detection— PE enables systematic exploration of diverse execution paths, expanding the scope of error detection within the source code.

\section{{\tool} Algorithm}

\subsubsection*{\bf Predictive Coverage-Guided Testing Framework Overview}
\label{sec:framework}

Fig.~\ref{fig:fuzzwise} shows the overall workflow with a multi-agent framework of two LLMs: 1) the LLM-based Test Case Generator (TCG), and 2) the LLM-based predictive executor {\model}. Both work collaboratively in an
iterative fashion in two phases to generate tests to
detect runtime~errors.
Initially, the LLM-based TCG analyzes the given code snippet and generates a test case. Starting with no coverage, each generated test case is fed into {\model} to predict code coverage. 
If total coverage keeps increasing, the TCG is prompted to both improve coverage in subsequent iterations, and detect runtime errors if possible. Once 100\% coverage is reached or the total coverage does not increase for a certain time limit, {\tool} instructs the TCG to generate test cases aimed only at triggering runtime errors. Each generated test case is fed into {\model} for coverage prediction and error detection. The process continues until the time limit.





\definecolor{gray}{rgb}{0.5,0.5,0.5}
\definecolor{mauve}{RGB}{127,0,145}
\definecolor{lightgray}{gray}{0.97}


\begin{figure}[t]
\lstset{
  language=Java, caption=, float, label={algo}, morekeywords={do, foreach, function, equals, and, or, in, not in, extend, append, then}, mathescape=true, escapechar=|,backgroundcolor=\color{white},
numbersep=-4pt,
numberstyle=\tiny\color{gray},
keywordstyle=\color{mauve},
    basicstyle=\scriptsize\sffamily  
}
\begin{lstlisting}[]
    $Test\_case$ = $\emptyset$
    $Cumulative\_coverage$ = $\emptyset$
    $Stop$ = False
    $Predicted\_coverage$ = $\emptyset$
    while (not $Stop$):
        if ($Cumulative\_coverage$ == 100% or not increasing in a time limit):
            $T\_case$ = TCG_LLM($error\_triggering\_prompt$)
        else:
            $T\_case$ = TCG_LLM($coverage\_increasing\_and\_error\_triggering\_prompt$)
        DuplicationRemoval($Test\_case$)
        $[Predicted\_coverage, RTEs]$=PE_LLM($T\_case$, $PUT$, $prompt$)
        Report runtime errors $RTEs$ (if any)
        $Cumulative\_coverage$ = update($Cumulative\_coverage$, $Predicted\_coverage$)
    if (time_limit_reached):
        $Stop$ = True
\end{lstlisting}
\vspace{-15pt}
\caption{{\tool} Algorithm}
\label{fig:algo}
\end{figure}


\subsubsection*{\bf Detailed Algorithm}


Fig.~\ref{fig:algo} presents the pseudo-code outlining {\tool}’ workflow for test case generation and execution-free runtime error detection. The algorithm begins by initializing essential variables: $Test\_case$ is an empty set to store generated test cases, $Cumulative\_coverage$ is initialized to track the total coverage of the {\em Program Under Test} 
 (PUT) after every iteration, $Stop$ is set to False to control termination, and $Predicted\_coverage$ is initialized to store the predicted code coverage (lines 1–4). The process iterates until a predefined time limit is reached (lines 5–15).

At each iteration, the algorithm dynamically selects between two test case generation strategies based on the overall code coverage status. If coverage remains incomplete (lines 8--9, Fig.~\ref{fig:algo}), the test case generation LLM is prompted with both strategies shown in Fig.~\ref{fig:cov-prompt} and Fig.~\ref{fig:exception-prompt}, i.e., generating test cases that target uncovered portions of the code to improve overall coverage and detect the runtime errors at the same time. Once total coverage reaches 100\%, meaning all statements have been covered by the existing test suite, or the coverage does not increase after a time limit (lines 6--7, Fig.~\ref{fig:algo}), the algorithm shifts focus to runtime error detection. At this stage, TCG\_LLM is prompted with a single strategy (Fig.~\ref{fig:exception-prompt}), creating test cases aimed to trigger potential runtime exceptions.


To maintain efficiency and prevent the redundant processing, the algorithm removes duplicate test cases (line 10) before passing a unique one to {\model}, {\tool}' predictive executor. {\model}\_LLM predicts the test case's coverage on the PUT and assesses its likelihood of triggering a runtime error, reporting any detected issues (lines 11--12). {\tool} then updates $Cumulative\_coverage$ based on $Predicted\_coverage$ (line 13), ensuring accurate tracking as new test cases are generated. This iterative process continues until the predefined time limit is reached (lines 14--15). 

\begin{figure}[t]
  \centering
	\lstset{
		numbers=left,
		numberstyle= \tiny,
		keywordstyle= \color{blue!70},
		commentstyle= \color{red!50!green!50!blue!50},
		frame=shadowbox,
		rulesepcolor= \color{red!20!green!20!blue!20} ,
		xleftmargin=1.5em,xrightmargin=0em, aboveskip=1em,
		framexleftmargin=1.5em,
                numbersep= 5pt,
		language=Java,
    basicstyle=\scriptsize\ttfamily,
    numberstyle=\scriptsize\ttfamily,
    emphstyle=\bfseries,
                moredelim=**[is][\color{red}]{@}{@},
		escapeinside= {(*@}{@*)}
	}
\begin{lstlisting}[caption = ]
(*@{\color{orange}{PROMPT} Coverage Increasing Test Case Generation@*)
Generate a test case for a program to cover uncovered lines of code
denoted by '!'. Provide only the test input without explanations. Consider various conditions, edge cases, and typical use cases.
Ensure the test case input is in the following format:
Test Case Input:
<input 1>
<input 2>...
\end{lstlisting}
\vspace{-15pt}
\caption{Portion of prompt to generate test cases to `increase coverage'}
\label{fig:cov-prompt}
\end{figure}

\begin{figure}[t]
	\lstset{
		numbers=left,
		numberstyle= \tiny,
		keywordstyle= \color{blue!70},
		commentstyle= \color{red!50!green!50!blue!50},
		frame=shadowbox,
		rulesepcolor= \color{red!20!green!20!blue!20} ,
		xleftmargin=1.5em,xrightmargin=0em, aboveskip=1em,
		framexleftmargin=1.5em,
                numbersep= 5pt,
		language=Java,
    basicstyle=\scriptsize\ttfamily,
    numberstyle=\scriptsize\ttfamily,
    emphstyle=\bfseries,
                moredelim=**[is][\color{red}]{@}{@},
		escapeinside= {(*@}{@*)}
	}
\hspace{4pt}
\begin{lstlisting}[caption = ]
(*@{\color{orange}{PROMPT} Runtime-error-detection Test Case Generation@*)
Generate a test case without providing an explanation for to raise the following scenarios in a Java program:
InputMismatchException: Provide an input value that whose data type is different than the one specified. 
ArithmeticException: Test cases that could raise arithmetic exceptions include division by zero, overflow, underflow, and attempts to perform invalid operations, e.g., taking the square root of a negative number.
NullPointerException: Create a scenario where a variable is explicitly set to null before usage.
NumberFormatException: A value that cannot be parsed to the expected data type, e.g., a non-numeric string.
ArrayIndexOutOfBoundsException or IndexOutOfBoundsException: Design input values leading to accessing array or list indices beyond their bounds.
(Other types of runtime errors and exceptions)
Ensure the test case input is in the following format:
Test Case Input:
<input 1>
<input 2>...
Generate a test case without any explanation for the below Java program:
\end{lstlisting}
\vspace{-15pt}
\caption{Portion of prompt to generate test cases intended to `catch runtime errors'}
\label{fig:exception-prompt}
\end{figure}

\section{The Multi-Agent Architecture of {\tool}}
\label{sec:llm}





The test generation and predictive execution are orchestrated through a structured interaction between multiple LLM agents. Specifically, test case generation is handled via two prompting phases directed at $TCG\_LLM$ (lines 7--9, Fig.~\ref{fig:algo}), while ${\model}\_LLM$ (line 10) performs predictive execution.


\subsection{The Test Case Generation Module}


The Test Case Generation (TCG) module formulates test cases based on the current coverage status of the test suite. Within the iterative test generation loop (line 5, Fig.~\ref{fig:algo}), two distinct prompting strategies are employed. The first phase aims to increase the overall code coverage and to detect errors, while the second strategy focuses only on identifying unhandled runtime errors/exceptions.

\subsubsection{{\bf [Phase 1]} Prompting for Test Case Generation to Increase Coverage and Detect Runtime Errors}

As {\tool} functions as a coverage-guided exploratory testing mechanism,
increasing coverage stands as a primary objective. If the overall coverage of the PUT is not 100\%, or newly generated test cases fail to expand coverage, the LLM is explicitly prompted to target unexplored/uncovered branches in the next cycle (Fig.~\ref{fig:cov-prompt}) and detect the runtime errors (Fig.~\ref{fig:exception-prompt}). 
The model is also instructed to generate test cases across a range of input categories, including boundary values (minimum, maximum), zero inputs, and both positive and negative cases, ensuring comprehensive test diversity.




\subsubsection{{\bf [Phase 2]} Prompting for Test Case Generation towards Runtime Error Detection}


Once full coverage is achieved or the coverage does not increase in a certain time limit, the focus shifts toward detecting runtime exceptions. In this phase, the LLM is guided to generate test cases specifically designed to trigger and expose potential runtime errors. The structure of this prompt (Fig.~\ref{fig:exception-prompt}) emphasizes predictive reasoning about vulnerable sections in code, ensuring that the generated test cases systematically explore conditions that might lead to unexpected failures.

\subsubsection*{\bf Note:} To maintain consistency and effectiveness, the output format remains standardized across both test generation strategies. This ensures that the transition between phases is seamless, facilitating efficient test case synthesis while maximizing the likelihood of error detection.




\subsection{LLM-based Code Coverage Prediction}
\label{sec:codepilot}




Unlike conventional approaches that solely estimate the coverage of a given test case, {\model} employs a structured reasoning framework to analyze execution paths, identify potential unhandled exceptions, and provide an explanation for their occurrence. 

To achieve this, {\model} provides additional constraints that direct the LLM to consider failure-inducing conditions. The model not only predicts the set of statements likely to be executed for a given test input, but also evaluates their susceptibility to errors such as \code{DivisionByZero}, \code{NullPointerException}, \code{OutOfBoundExceptions}, etc. This dual-faceted reasoning process ensures that the output of {\model} includes (1) the predicted code coverage, (2) a list of potential runtime errors associated with the test case, and (3) a step by step reasoning detailed why these errors might occur under specific input conditions. Thus, {\model} helps {\tool} in reducing reliance on direct execution while simultaneously increasing the efficiency of identifying critical software flaws.




\section{Empirical Evaluation}
\label{sec:eval}

For evaluation, we seek to answer the following questions:

\noindent {\bf RQ1. [Effectiveness and Efficiency of Runtime Error Detection on Java Code].} How does the performance of {\tool} in runtime error detection compared to traditional and ML fuzzers when applied to complete code?

\noindent {\bf RQ2. [Efficiency of Test Case Generation].} If each unique exception represents a distinct point of unexpected behavior within the execution space, how swiftly does {\tool} transition between these instances during testing?



\noindent {\bf RQ3. [Effectiveness of Runtime Error Detection on Python].} How well does {\tool} detect runtime errors on Python code?

\noindent {\bf RQ4. [Ablation Study].} How do components in {\tool} contribute to its performance?

\noindent {\bf RQ5. [Static Detection and Recommendation of Exception Handling Errors on S/O Code Snippets].} 
How well does {\tool} detect and recommend exception handling errors in SO code?

\section{Runtime Error Detection on Java Code Snippets (RQ1)}
\label{sec:rq2-r2}



\subsection{Baselines, Dataset, Procedure, and Metrics}

\subsubsection{Baselines}
We chose Jazzer~\cite{jazzer} and LLM-based
Fuzz4All ~\cite{xia2024fuzz4all} as baselines. Jazzer is a
mutation-based coverage-guided fuzzer for Java programs.
Fuzz4All is a fuzzer that leverages LLMs to support various
languages. The common aspect between the baselines is
that like all existing fuzzers, they execute each generated test
case on the PUT to detect errors and compute coverage.

\subsubsection{Dataset} 
We used the FixEval dataset~\cite{haque2023fixeval}, a benchmark
comprising submissions to competitive programming
challenges in AtCoder~\cite{atcoder} and Aizu Online
Judge~\cite{aizu}. 
We selected 200 diverse Java code snippets from distinct problem statements in this dataset (100 complete and 100 incomplete code snippets). The chosen
code examples varied across multiple dimensions, including code length,
branches, and the presence of loops.

\subsubsection{Procedure} We used GPT-3.5 (for cost efficiency) for TCG, and GPT-4o for PE.
All tools were allowed to run for 5 minutes per program. For Jazzer and Fuzz4All, this period
encompasses the fuzzing process, excluding the
compilation and building phases to ensure fair comparison because
{\tool} does not require those phases. For {\tool}, the
5-minute limit comprises all the steps within the framework
(Section~\ref{sec:framework}) including the code coverage and runtime error prediction. Upon reaching
the time limit for each data sample, the models were stopped and the results were analyzed.
In experiments of RQ1 and RQ5 on runtime error detection for incomplete code from StackOverflow and FixEval datasets, we leveraged the corresponding complete versions of the incomplete code from prior works~\cite{crispe2025,cai2024programming,nguyen2020coderecommendation}. For correctness verification, we executed the corresponding complete code with the generated test cases to verify whether a predicted runtime error occurs.  


\subsubsection{Evaluation Metrics}
We used two metrics for evaluating the effectiveness of each tool: {\em Error
Trigger Rate ({\bf ETR})} and {\em Bug Discovery Rate ({\bf BDR}).} {\em The Exception Trigger
Rate} is defined as the ratio between the number of generated test
cases that are effective in triggering exceptions to the total number of
generated test cases. A higher ratio indicates more effective
in finding runtime errors. {\em The Bug Discovery Rate} measures how many
unique runtime errors are found in a period of time.



\subsection{Empirical Results}

\begin{figure}[t]
\begin{center}
\includegraphics[width=3.3in]{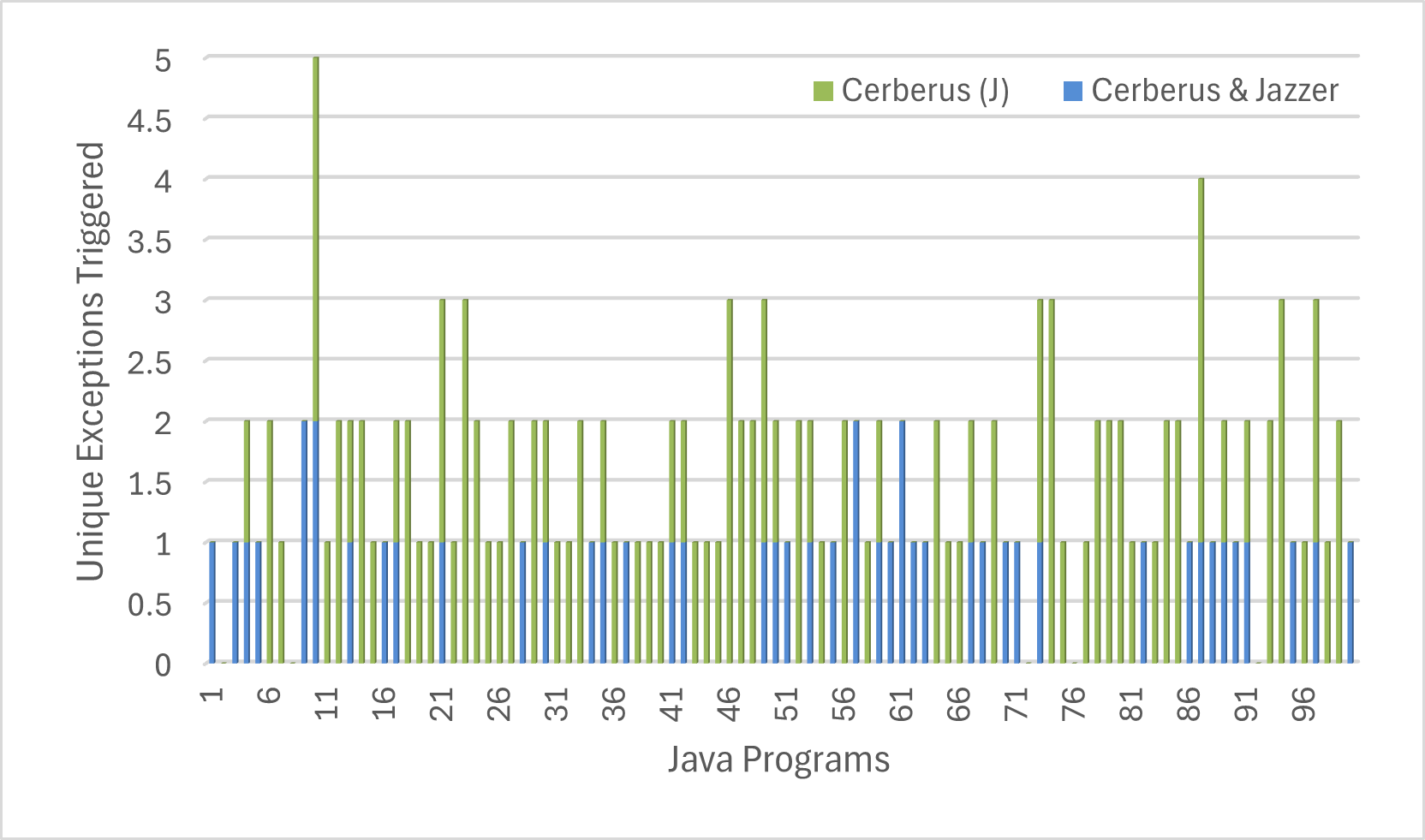}
\vspace{-9pt}
\caption{Runtime Error Detection: {\tool} vs Jazzer (RQ1)}
\label{fig:rq1-cerberus-jazzer}
\end{center}
\end{figure}

\begin{figure}[t]
\begin{center}
\includegraphics[width=3.3in]{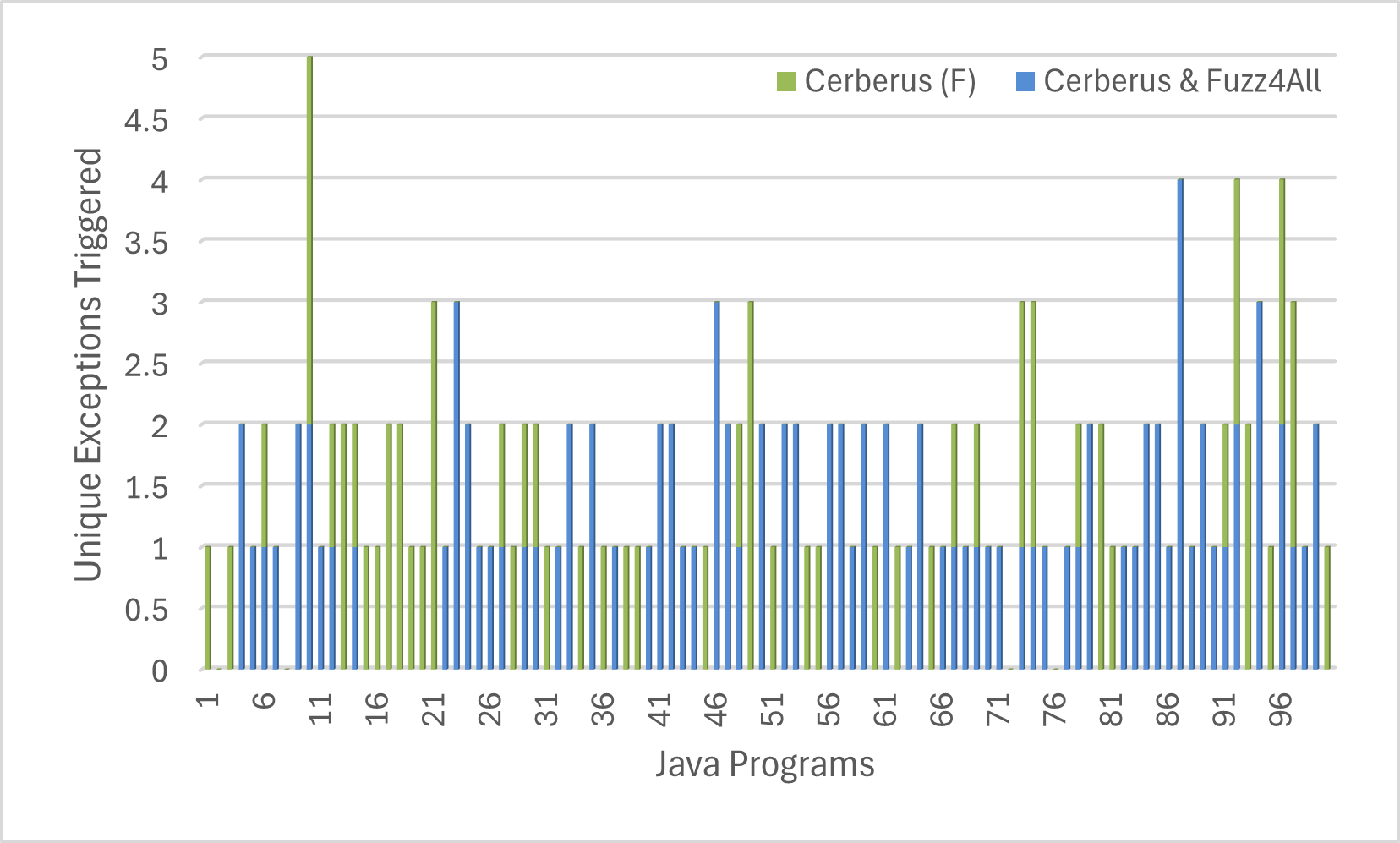}
\vspace{-9pt}
\caption{RT Error Detection: {\tool} vs Fuzz4All (RQ1)}
\label{fig:rq1-cerberus-fuzz4all}
\end{center}
\end{figure}

\subsubsection{Runtime Error Detection Effectiveness}

{\em The Bug Detection Rates (BDR)} for all approaches are depicted in
Fig.~\ref{fig:rq1-cerberus-jazzer} and
Fig.~\ref{fig:rq1-cerberus-fuzz4all}. Each value on the Y-axis
indicates the number of unique runtime exceptions that a fuzzer was able
to trigger/detect within a 5-minute period, with the points plotted for every
program in our dataset (X-axis). In each figure, the blue bars
represent programs for which both baseline and {\tool} detect the same
number of exceptions, while the green bars denote the additional
exceptions that {\tool} predicts correctly but the baseline does
not. As seen, {\tool} consistently outperforms both baselines in
detecting more unique runtime errors/exceptions. {\tool}'s Bug
Detection Rate is either equal to or higher than both baselines for
any Java program in the dataset. In
Fig.~\ref{fig:rq1-cerberus-jazzer}, while Jazzer can detect a maximum
of one runtime error within the time limit, {\tool} can correctly
identify up to five runtime errors. Moreover, Jazzer fails to trigger
any exceptions in 55 cases (indicated by solid green or no bar from
X-axis), whereas {\tool} only fails in 5 programs.  Moreover, in 
Fig.~\ref{fig:rq1-cerberus-fuzz4all}, {\tool} also
outperforms Fuzz4All. Fuzz4All is able to trigger a maximum of four
runtime errors, while {\tool} detects five. Fuzz4All fails to trigger
any exceptions for 33 programs, compared to only 5 programs for {\tool}.

As seen in Table~\ref{tab:rq1-javacerberusprediction}, a precision
value of 0.7205 indicates that it correctly detect the specific
runtime errors in approximately 72.05\% of the instances.
A recall value of 0.5617 implies that it is able to detect
approximately 56.17\% of all actual runtime errors present in the code
snippets in the sampled dataset. 
This performance advantage can be attributed to the implementation of {\model}, which uses a planning-based approach to predict code coverage and detect runtime errors. The use of step-by-step reasoning has a positive impact on the accuracy of prediction as it allows for a more thorough analysis of the execution paths within the code, enabling the model to systematically evaluate each step and its implications on coverage and potential errors. By breaking down the process into manageable steps, {\model} can better account for complex interactions within the code, reducing the likelihood of oversights and enhancing overall detection accuracy.

\begin{table}[t]
  \centering
  \footnotesize
  \caption{Error detection effectiveness for complete code in Java (RQ1)}
  \vspace{-6pt}
  \label{tab:rq1-javacerberusprediction}
  \renewcommand{\arraystretch}{1.4}
  \begin{tabular}{l|c|c|c}
    \hline
    \textbf{Model} & \multicolumn{3}{c}{\textbf{Evaluation Metrics}} \\ \cline{2-4}
    & \textbf{Precision} & \textbf{Recall} & \textbf{F1-Score} \\ \hline
    {\tool} & {0.72} & {0.56} & {0.63} \\ \hline
  \end{tabular}
\end{table}

\begin{table}[t]
  \centering
  \small
  \caption{Error detection effectiveness for incomplete code in Java (RQ1)}
  \vspace{-6pt}
  \label{tab:rq1-incomplete}
  \renewcommand{\arraystretch}{1.4}
  \begin{tabular}{l|c|c|c}
    \hline
    \textbf{Model} & \multicolumn{3}{c}{\textbf{Evaluation Metrics}} \\ \cline{2-4}
    & \textbf{Precision} & \textbf{Recall} & \textbf{F1-Score} \\ \hline
    {\tool} & 0.58 & 0.41 & 0.48 \\ \hline
  \end{tabular}
\end{table}

Table~\ref{tab:rq1-incomplete} displays the result of {\tool} on incomplete code. A precision
value of 0.58 indicates that it correctly detects the specific
runtime errors in approximately 6 out of 10 instances. A recall value of 0.41 implies that it is able to detect approximately 4 out of 10 actual runtime errors present in the code
snippets in the sampled dataset. Compared with Table~\ref{tab:rq1-javacerberusprediction}, we can observe that {\tool}'s performance on the complete code is better than that on the incomplete one. This is reasonable because the incomplete code snippets in our dataset often miss `\code{import}' statements, variable declarations, and type declarations. This incompleteness hinders the understanding for LLMs more than the complete code. 

\begin{table}[t]
  \centering
  \small
  \caption{Error detection efficiency for Java code (RQ1)}
  \vspace{-6pt}
  \label{tab:rq1fuzzer}
  \renewcommand{\arraystretch}{1.2} 
  \begin{tabular}{l|c|c|c}
    \hline
    \textbf{Fuzzer} & \multicolumn{2}{c|}{\textbf{Test Cases (Average across dataset)}} & \textbf{ETR (\%)} \\ \cline{2-3}
    & \textbf{Generated} & \textbf{Effective} & \\ \hline
    Jazzer & 3,123,232 & 300,830 & 9.63 \\ \hline
    Fuzz4All & 84 &15 & 17.85 \\ \hline
    \textbf{{\tool}} & \textbf{9} & \textbf{3} & \textbf{33.33} \\ \hline
  \end{tabular}
\end{table}

\subsubsection{Runtime Error Detection Efficiency}

As seen in Table~\ref{tab:rq1fuzzer}, {\tool} achieves the highest ETR
at 33.33\%. That is,~out of an average of 9 test cases generated by
the Test Case Generator (TCG\_LLM) for a program within the time
limit, {\model} accurately predicts 3 of them as error-triggering
test cases. Notably, this metric includes only the predictions where
the type of exception must be correctly identified, ensuring 100\%
accuracy. When compared to the baselines, {\tool} shows a good 
improvement, performing 2.46X and 86.72\% better
than Jazzer and Fuzz4All respectively.

Both {\tool} and Fuzz4All leverage GPT in generating ({\tool}) and mutating (Fuzz4All) a variety of
test cases, rather than the traditional mutation used by Jazzer. Thus,
the number of the generated test cases are much smaller than that for
Jazzer. {\tool} is also more efficient in error detection than
Fuzz4All with almost double ETR. We could attribute higher efficiency
of {\tool} over Fuzz4All due to their different architectures
(see Section~\ref{sec:coverage-plateau}).
Note that even with a higher number of effective test cases triggering
errors, the unique number of detected runtime errors from Fuzz4All
 is still smaller than that of
{\tool}. Moreover, Jazzer and Fuzz4All require
a meticulous setup process and a significant number of additional
dependencies, which \tool{} minimizes.


The significantly higher ETR indicates that {\tool} is adept
at predicting whether a test cases is effective in
triggering an exception. The higher BDR emphasizes its capability to trigger a broader range of unique errors within the time limit, outperforming the baselines.



\subsubsection{Coverage Plateau}
\label{sec:coverage-plateau}
  
When a testing tool is initially deployed, it typically identifies numerous new execution paths and achieves substantial code coverage rapidly.
As the process progresses, the discovery of new paths diminishes. The remaining unexplored paths are more complex, necessitating specific and potentially rare input conditions to be activated. Consequently, the coverage growth flattens out, resulting in a phenomenon called {\em ``coverage plateau''}~\cite{10172800}.
%


\begin{figure}[t]
\begin{center}
\includegraphics[width=3.1in]{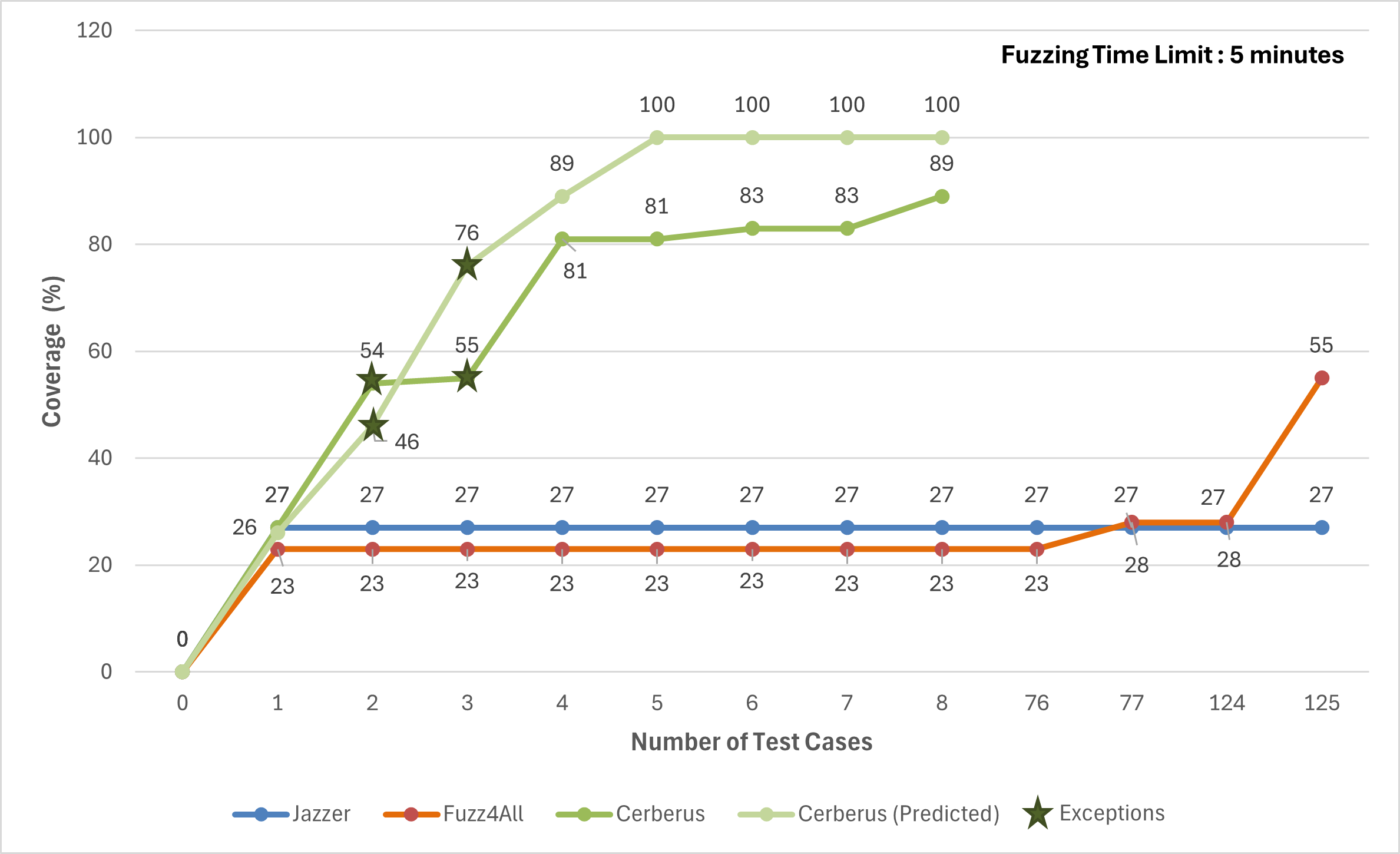}
\vspace{-8pt}
\caption{Coverage Plateau: {\tool} vs Baselines (RQ1)}
\label{fig:rq1-coverage-plateau}
\end{center}
\end{figure}

We aim to illustrate the coverage plateau from three approaches under study
via a medium-length example. As seen in Fig.~\ref{fig:rq1-coverage-plateau},
Jazzer quickly reaches a plateau at 27\% coverage after the initial
test cases and maintains this level throughout the period of 5
minutes. This stagnation indicates Jazzer's difficulty in uncovering
new code paths beyond the initial discoveries. Fuzz4All attains an
early plateau at 23\% coverage, maintaining this level for most of 
test cases. However, a sudden spike occurs at test cases \#124 and \#125,
increasing coverage to 28\% and then to 55\%. This late
discovery of a significant path suggests that Fuzz4All experiences
prolonged coverage plateaus but can occasionally find substantial new
paths.

In contrast, {\tool} exhibits a different pattern. Starting with 0\% coverage, it rapidly increases to 12\% after the first test case and quickly progresses to 54\% and then 55\% with the first few test cases. It experiences brief intervals~of small plateaus, spanning two test cases each, at around 81\% and 83\% coverage for test cases \#4 to \#5 and \#6 to \#7. Ultimately, it achieves 89\% coverage by test case \#8, showing fast coverage improvement with short periods of plateau.


{\tool} outperforms Jazzer and Fuzz4All due to its ability to quickly generate relevant test cases targeting untested code paths leads to rapid initial coverage gains and brief plateaus. By achieving 89\% coverage by test case \#8, {\tool} showcases its effectiveness in both increasing code coverage and detecting runtime errors efficiently. 

In total, within 8 test cases, {\tool} reaches the maximum coverage of 89\% with a total plateau of 2 iterations. For Fuzz4All, in 125 generated test cases, it reaches the coverage of 55\% with a total plateau of 122 iterations.
For Jazzer, within 125 iterations, it reaches the maximum coverage of 27\%m with a long plateau of 124 iterations.

\subsubsection{Predicted Coverage versus Actual Coverage} When compared to the ground truth, i.e., the actual running of the test cases produced by {\tool} (the darker green line in Fig.~\ref{fig:rq1-coverage-plateau}), the predicted coverage shows variation across all test cases but with minimal differences. {\tool} (Predicted) is 100\% accurate in error prediction (denoted by green stars on both {\tool}'s lines) by each test case. CODAMOSA~\cite{10172800} was not tested since it generates only unit tests.



\section{Efficiency of Test Case Generation (RQ2)}
\label{sec:rq3}

\subsection{Procedure and Metrics}




In RQ1, we showed that {\tool} can detect more runtime errors
with a smaller number of test cases.
In this RQ2, we aim to further evaluate its efficiency in
transitioning the detection from one to another runtime error. In
a coverage-guided testing framework, encountering a bug or runtime error indicates
that a tool has discovered a path in the program's execution space
leading to unexpected behavior. This represents a distinct {\em
  ``local minimum''} of undesirable behavior. An ideal tool
minimizes the number of test cases needed to identify the next local
minimum. Fewer test cases indicate that the tool is more efficient and faster
in exploring new paths in the execution space, leading to
its efficiency in detecting runtime errors.


Specifically, we investigate {\em the number of test cases needed to
  detect the next unique runtime error}. 
Using the results from RQ1, we determine the number of test cases
required for each tool to find the next runtime error. This
metric is averaged across the dataset.

\begin{figure}[t]
\begin{center}
\includegraphics[width=3.3in]{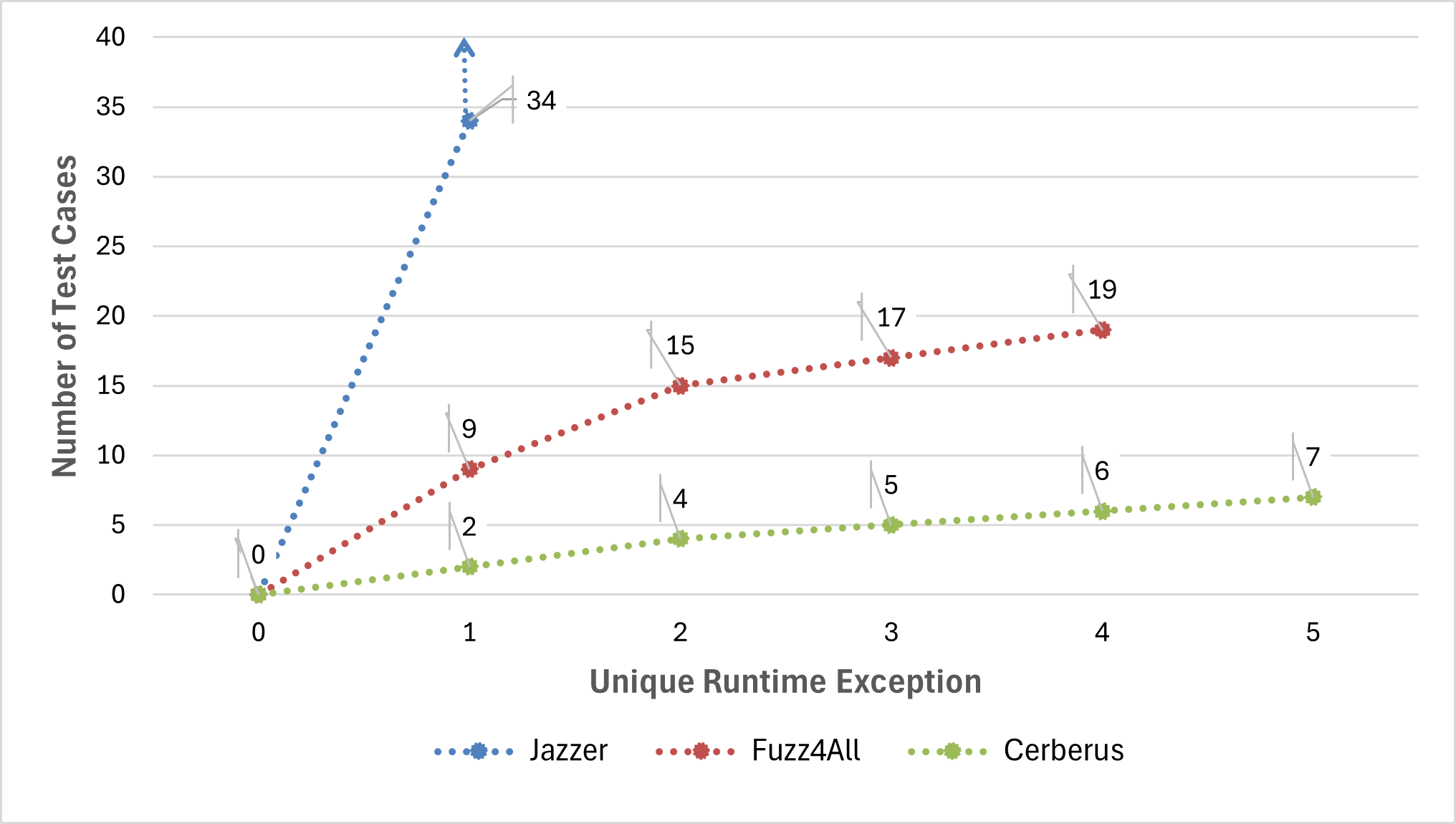}
\vspace{-9pt}
\caption{Efficiency of Test Case Generation (RQ2)}
\label{fig:rq3_local_minima}
\end{center}
\end{figure}

\subsection{Empirical Results}

Fig.~\ref{fig:rq3_local_minima}  displays the number of test cases needed for the tools to identify unique runtime errors.
Jazzer requires an average of 34 test cases to detect the first
error in a code snippet. It then takes over a
million test cases on average to detect the next error. The
quantity of test cases required is shown in the blue line (Fig.~\ref{fig:rq3_local_minima}).


Fuzz4All shows moderate efficiency. It requires fewer test
cases than Jazzer for each successive detected error. As seen, Fuzz4All (red line) generates an
average of {\bf 9} test cases before triggering its first
error. The second one is caught after an average of {\bf
  6} additional test cases, while the third and fourth previously
unique exceptions are caught after generating an additional
{\bf 2} test cases each.

{\tool} exhibits the highest efficiency, requiring the fewest test
cases to detect new unique runtime errors. As seen in
Fig.~\ref{fig:rq3_local_minima}, it (green line) generates an
average of only {\bf 2} test cases to trigger its first exception and
continues to require {\em 2 or fewer test cases} between each unique
exception detected. That is, following the first
exception, in a fixed number of subsequent test cases generated,
it has the highest likelihood of catching other unique errors.




We can attribute {\tool}'s faster transition between local minima to
its test case generation strategy. Jazzer uses
blind mutation and does not utilize the code
understanding capabilities of LLMs. On the other hand, Fuzz4All uses an LLM to select and
perform mutations for generating new test cases. However, {\em Fuzz4All
still relies on the mutation strategy to generate its next set of test cases in each iteration}. In contrast, {\bf {\tool} capitalizes on the advanced code comprehension capabilities of LLMs, to generate test cases that are strategically focused on uncovering specific runtime errors}. This eliminates the need for traditional or LLM-based mutation strategies, allowing {\tool} to efficiently explore the codebase and quickly identify new runtime errors. By integrating predictive coverage analysis with targeted test case generation, it enhances overall error detection.


\section{Runtime Error Detection on Python (RQ3)}
\label{sec:rq4}

The goal of this study is to evaluate {\tool}'s effectiveness and its utility in
fuzzing Python code to detect runtime errors without actual execution.


\begin{table}[t]
  \centering
  \small
  \caption{Effectiveness in Runtime Error Detection for (In)complete Python Code Snippets (RQ3)}
  \vspace{-6pt}
  \label{tab:rq4pythoncode}
  \renewcommand{\arraystretch}{1.65}
  \begin{tabular}{l|c|c|c}
    \hline
    \textbf{Model} & \multicolumn{3}{c}{\textbf{Evaluation Metrics}} \\ \cline{2-4}
    & \textbf{Precision} & \textbf{Recall} & \textbf{F1-Score} \\ \hline
    {\tool} & {0.7429} & {0.7981} & {0.7695} \\ \hline
  \end{tabular}
\end{table}

\subsection{Dataset, Procedure, and Metrics}

\subsubsection{Dataset}

A subset of the FixEval dataset, consisting of 100 Python code snippets, was used for this study. Among these, {\em 25 had incomplete code} leading to \code{NameErrors} from missing imports, 25 had \code{SyntaxErrors}, and {\em the remaining 50 were complete}. All samples were mixed and subjected to fuzzing to trigger runtime errors, stratified and randomly selected based on complexity and length, focusing on medium to large snippets averaging 100 or more statements.

\subsubsection{Procedure}

Consistent with the evaluation in RQ1, each
(in)\-complete code snippet was subjected to a test generation process using {\tool}
for 5 minutes. The predicted run-time errors were
later analyzed. The ground truth was established
by executing each generated test case on the original complete code or
the corresponding complete code snippets of the incomplete ones.


\subsubsection{Metrics}

To assess {\tool} in detecting runtime errors, we use three evaluation metrics: Precision, Recall, and
F1-Score. Precision is defined as the ratio of true positive
predictions to the total number of positive predictions (both true and
false positives). Conversely, Recall is the ratio of true positive
predictions to the total number of actual positives (true positives
and false negatives). The F1-Score is the harmonic mean of precision
and recall. 

\subsection{Empirical Results}

As seen in Table~\ref{tab:rq4pythoncode}, a precision value of 0.7429
indicates that it correctly detect the specific runtime errors in
approximately 74.29\% of the instances. This high precision suggests
that {\tool} is quite accurate in identifying true positives,
resulting in relatively few false positives. Similarly, a recall value
of 0.7981 implies that it is able to detect approximately 79.81\% of
all actual runtime errors present in the code snippets in the sampled
dataset. This high recall rate indicates that it is effective in
identifying a large portion of actual runtime errors, leading to
relatively few false negatives. An F1-Score of 0.7695 indicates a good
balance between precision and recall. This relatively high F1-Score
shows that {\tool} performs well in both accurately
identifying runtime errors and ensuring that most actual errors are
detected. Among 50 incomplete code snippets in Python, {\tool} 
detected 43 of them with 100\% accuracy in all the runtime errors in
those snippets.


The results indicate that {\tool} is highly effective in
detecting runtime errors in Python code for both (in)complete code. 
Generally speaking, {\tool} performs better on
detecting runtime errors for Python (in)complete code snippets.
Importantly, this demonstrates that {\tool} can be applicable to more
than one programming languages due to its LLM-based mechanism.


\section{Ablation Study (RQ4)}
\label{sec:rq1}




\subsection{Baselines, Dataset and Metrics}

\subsubsection{Baselines} 

The baseline models include

\underline{Baseline 1}. {\em\textbf{Single-agent Basic}}: this
baseline uses an LLM with a {\bf single vanilla prompt} to directly derive error-triggering inputs and detect runtime errors (Section~\ref{sec:motiv}).
This represents the non-modular, LLM-based approach {\em without two-phase, feedback loop and multi-agents}.

\underline{Baseline 2}. {\em\textbf{Multi-agent Basic}}: this baseline has a
multi-agent architecture with two LLMs: one for test case generation
and the other for coverage prediction. However, the first LLM generates the test cases and the second 
operates with one-shot prompting for code coverage prediction without feedback loop. This represents 
the LLM-based {\bf multi-agent} approach {\bf without the feedback loop}. 

\underline{Baseline 3}. {\em\textbf{Multi-agent Feedback}}:
This variant represents the {\bf direct simulation of the traditional, coverage-guided testing framework with a single phase} in which the first LLM performs test case generation (i.e., LLM-based mutation) and the second LLM for predictive execution predicts the code coverage. This variant is similar to Multi-agent Basic, however, it is enhanced it with a feedback loop from the second LLM to the first one.
Compared with {\tool}, Multi-agent Feedback merges two goals  (generating test cases for both coverage-increasing and error detection) into one phase for the first LLM with {\model} used for code coverage prediction for the generated test cases. It does {\em not use the two-phase feedback and dual-prompting}.

\subsubsection{Dataset and Metrics}

The dataset used for comparison is a carefully curated subset of FixEval, specifically containing 100 unique code snippets that exhibit runtime errors. These samples contain medium to large programs with varying levels of complexity for a diverse range of scenarios and are different than the ones in RQ1. We use the same metrics as in RQ1 (Precision, Recall, F1).

\subsection{Empirical Results}

\begin{table}[t]
  \centering
  \footnotesize
  \caption{Ablation Study on Model Architecture (RQ4)}
  \vspace{-6pt}
  \label{tab:rq1coverageeval}
  \renewcommand{\arraystretch}{1.3} 
  \begin{tabular}{l|c|c|c}
    \hline
    \textbf{Model Architecture} & \multicolumn{3}{c}{\textbf{Evaluation Metrics}} \\ \cline{2-4}
    & \textbf{Precision} & \textbf{Recall} & \textbf{F1-Score} \\ \hline
    Single-Agent Basic & 0.30 & 0.30 & 0.30 \\ \hline
    Multi-Agent Basic & 0.49 & 0.36 & 0.41 \\ \hline
    Multi-Agent Feedback & 0.52 & 0.38 & 0.44 \\ \hline
    \textbf{{\tool}} & \textbf{0.72} & \textbf{0.56} & \textbf{0.63} \\ \hline
  \end{tabular}
\end{table}




As seen in Table~\ref{tab:rq1coverageeval}, {\tool} achieves the
highest accuracy in~runtime error detection. With 58\% precision,
almost 6 out of 10 detected runtime errors are actual true. The recall
of 41\% indicates that it can detect 4 out of 10 possible runtime
errors in the ground truth.

\underline{First}, the single-agent basic model has lowest performance. The LLM
may inaccurately predict runtime errors, resulting in false positives. Even when errors are correctly identified, the generated inputs may fail to trigger them, leading to false negatives.

\underline{Second}, {\tool} relatively improves over the Multi-agent Basic by 18.4\% in precision, 13.9\% in recall, and 17\% in F1-score. The result shows that the division of responsibilities is helpful in which one LLM is dedicated to test case generation and another one to coverage prediction and error detection. However, the {\em absence of the feedback loop means that the coverage improvements achieved by the coverage prediction LLM are not fed back to guide the subsequent iteration}. Thus, the test
cases generated solely prioritize increasing code coverage without the necessary adjustments to specifically target exception-triggering scenarios. As a result, the Multi-Agent Basic architecture often fails to produce test cases that effectively trigger exceptions, as its focus remains on broad coverage rather than targeted exception and runtime error detection.



Lastly, {\bf {\tool} improves relatively over the Multi-Agent Feedback
model (i.e., direct simulation of coverage-guided testing framework)} with single prompt merging two goals 
by 11.5\% in precision, 7.9\% in recall, and 10\% in F1-score.
The Multi-Agent Feedback baseline integrates a feedback loop for
iterative test case refinement, similar to {\tool}, allowing the
system to iteratively refine test case generation based on previous
results. {\em With both goals of increasing code coverage and detecting runtime errors
in one phase, the test-generation LLM must have more burden in optimizing toward
both objectives}. In contrast, with the two-phase feedback, {\tool} allows
the LLM first focus on improving code coverage, which also enhances the opportunity of error detection. When the total coverage of the current test suite reaches 100\%, {\tool} guides
the LLM toward detecting runtime errors/exceptions, making the process more effective.

The relative improvement of {\tool} over this baseline can also be
attributed to its planning-based mechanism.
{\tool} employs a step-by-step reasoning process within the {\model} that allows for an understanding of code execution paths and their coverage.






\section{Detecting and Recommending Runtime Exception Handling on S/O Code (RQ5)}
\label{sec:so-experiment}

\subsection{Dataset, Procedure, and Metrics}



\subsubsection{Dataset} 

To evaluate {\tool} on real-world StackOverflow (SO) code, we focus on a type of runtime errors called {\em exception-related bugs}~\cite{cai2024programming}. We first utilized the mining tool from Zhang {\em et al.}~\cite{DBLP:conf/icse/0001YLK19} to identify SO code snippets and their corresponding adaptations in GitHub repositories. We assessed the {\em error-proneness} of a snippet by detecting exception-related errors as follows: A snippet is classified as containing a {\em runtime exception-related bug} if its original SO post lacks exception handling, but the adapted version includes explicit exception handling that the developer implemented to resolve the encountered error. Conversely, {\em non-buggy} snippets are those integrated into projects either unchanged or with modifications that did not introduce exception handling. Thus, from a total of 1,505 mined snippets with the above procedure, we curated and randomly selected a dataset of 100 Java snippets with 50 buggy and 50 non-buggy ones. Moreover, we chose another 100 non-buggy samples to test if {\tool} can recommend extra exceptions that were missed by the developer during adaptation.


\subsubsection{Procedure} 
We tested each StackOverflow code snippet with {\tool}, to assess its effectiveness in predicting and triggering runtime errors. The core idea is to analyze each snippet using {\tool} in order to evaluate whether it can do the following: 

1. {\bf Precisely identify the unhandled exceptions} that were later addressed in the adapted code (i.e., matching the ground truth).

2. {\bf Detect and recommend (additional) runtime errors} that were not addressed in the adapted code (ground truth), pointing out exceptions that were potentially missed during adaptation.


\subsubsection{Metrics} We used the following evaluation metrics: 



1. {\em True Positive (TP)} measures {\tool’s ability to correctly identify runtime exceptions that were originally unhandled in the S/O snippet but were explicitly handled in the ground truth.

2. {\em False Negative (FN)} for cases where {\tool} failed to detect an exception that was later handled in the ground truth. 

3. {\em True Negative (TN)} assesses {\tool}'s ability to correctly classify non-buggy snippets—i.e., those that were integrated into GitHub projects without any additional exception handling. 

4. {\em False Positive (FP)} refers to non-buggy instances being incorrectly predicted as an exception.



\subsection{Empirical Results}

\subsubsection{Detection of Runtime Exceptions}

As seen in Table~\ref{tab:rq5}, the TP of {\bf 63.15\%} indicates that {\tool} correctly predicted the exceptions handled in the adapted GitHub code for 36 out of 57 buggy cases. However, the False Negative of {\bf 36.85\%} reflects the {\bf 21 missed cases}, where {\tool} failed to detect the exceptions that were explicitly handled in the adapted code. Certain exceptions such as \code{UnsupportedEncodingException}, \code{XMLException}, and \code{NoSuchAlgorithmException} might only manifest under specific runtime conditions in which {\tool}' fuzzing strategies fail to generate test cases for. For non-buggy snippets, {\tool} achieved the True Negative of {\bf 84\%}, correctly identifying 42 out of 50 snippets as not requiring additional exception handling. The False Positive of {\bf 16\%} arises from {\bf 8 cases} where {\tool} incorrectly flagged an exception in otherwise non-buggy code. This could be due to over-generalization, where {\tool} identifies patterns resembling common exception-prone code structures  but misclassified them as actual errors (e.g., \code{UnsupportedEncodingException and IOException}). Another contributing factor might be the absence of explicit exception handling in the ground truth, meaning that while no handling was added in the GitHub adaptation, the original snippet could still be prone to runtime errors under other conditions, {\em leading {\tool} to flag potential but not-yet-verified issues}.

Overall, {\tool} achieves {\bf an accuracy of 72.89\%}, balancing its ability to effectively detect exception-related bugs in real-world SO code while needing to improve in minimizing false negatives and false positives. Next, let us present the cases where {\tool} correctly recommends more exception handling than the actual~code.





\begin{table}[t]
    \centering
    \small
\tabcolsep 2pt
\caption{Runtime Error Prediction on S/O code (RQ5)}
\vspace{-6pt}
\begin{tabular}{cc|cc|c}
\hline
\multirow{2}{*}{\textbf{}} &
  \multirow{2}{*}{\textbf{Actual}} &
  \multicolumn{2}{c|}{\textbf{Predicted}} &
  \multirow{2}{*}{\textbf{Accuracy}} \\ \cline{3-4}
 &
   &
  \multicolumn{1}{c|}{\textit{Buggy}} &
  \textit{Non-Buggy} &
   \\ \hline
 &
  \textit{Buggy} &
  \multicolumn{1}{c|}{\begin{tabular}[c]{@{}c@{}}TP = 36/57 (63.15\%)\end{tabular}} &
  \begin{tabular}[c]{@{}c@{}}FN = 21/57 (36.85\%)\end{tabular} &
   \\ \cline{2-4}

\multirow{-2}{*} &
  \textit{Non-Buggy} &
  \multicolumn{1}{c|}{\begin{tabular}[c]{@{}c@{}}FP = 8/50 ({16\%})\end{tabular}} &
  \begin{tabular}[c]{@{}c@{}}TN = 42/50 ({84\%})\end{tabular} &
  \multirow{-2}{*}{\textbf{72.89\%}} \\ \hline
\end{tabular}%
\label{tab:rq5}
\end{table}

\subsubsection{Recommendation of Runtime Exception Handling}

Beyond detecting previously handled exceptions, {\tool} also identified additional unhandled runtime exceptions, demonstrating its potential as a useful recommendation system for improving code robustness. Out of {\bf 100 StackOverflow code snippets (labeled as non-buggy)}, {\tool} accurately identified {\bf 79 samples that still contained unhandled exceptions}. All predictions were validated through execution, {\em confirming both the presence of the missed exceptions} and the effectiveness of the generated test case inputs in triggering each runtime bug. It is clear that even after adaptation into GitHub repositories, many runtime exceptions remained unhandled, exposing the code to potential runtime errors. This highlights human errors in manual exception handling, as developers may overlook certain edge cases. By proactively identifying and recommending these overlooked exceptions, {\tool} not only helps reveal runtime errors but also strengthens the overall reliability of the codebase, maximizing bug detection and address potential failures before~deployment. 

\paragraph{Dataset and Case studies} The complete dataset used for runtime error detection on real-world S/O code snippets, along with a detailed case study can be found in our project's website~\cite{cerberus-website}.




\section{Limitations and Threats to Validity}
\label{sec:limitations}

\indent {\em Limitations.} There are still rooms for improvement in {\tool}.
First, the inherent complexity and probabilistic nature of black box LLMs can lead to variations in behavior in specific edge cases. However, running the generated test cases remains a robust method for detecting runtime errors effectively. Second, while coverage prediction offers a useful estimation of code paths, its performance may vary when applied to previously unseen libraries. Third, using LLMs would incur a cost. For one code snippet, the detection time is 5 seconds on average, and the token consumption is 1,200 (input) and 6,500 (output) tokens. Finally, we aim to support much larger programs, but current focus is on runtime error detection in online (in)complete code snippets, which are typically a few hundred lines long.



{\em Threat to Validity}. We use only GPT-4o for our experiment. The results might vary for other LLMs.
They may be influenced by the specifics of the testing environment, including hardware and software configurations. 
Our FixEval dataset might not be representative. The results might vary on different datasets. 
The LLMs in our experiments might have {\em hallucinations}. We mitigate them by two strategies: 1) we set the temperature of zero to reduce that, and 2) for complete code, we could verify each generated test case through execution, confirming whether it triggers a runtime error. 
%
While the FixEval dataset predates GPT-4o, it only provides exception labels (e.g., \code{IndexOutOfBounds}) without the inputs that trigger them. Similarly, the SO benchmark includes buggy snippets with patches but not the corresponding error-inducing inputs. In Cerberus, the LLM must infer such inputs through execution reasoning, making data leakage unlikely.

To further address that, our ablation study (RQ4, Table~\ref{tab:rq1coverageeval}) compared {\tool} with vanilla prompting using GPT-4o (Single-agent basic). {\tool} significantly outperforms the baseline, indicating the gains stem from Cerberus itself, not data leakage.

\section{Related Work}
\label{sec:related}

Fuzzing has been extensively studied. Miller {\em et al.}~\cite{miller1995fuzz} conducted an empirical
study evaluating the reliability of UNIX utilities, pioneering the
concept of fuzzing by exploring the effectiveness of providing
unexpected or random data as input to uncover bugs.


As fuzzing progresses, efforts shifted to enhancing the process's
efficiency and effectiveness. {\em Coverage-guided
  fuzzing}~\cite{JQF,10.1145/3133956.3138820}
emerged as an advancement, addressing the need for a systematic
approach to explore the vast input space of software. By monitoring
code execution with each input and prioritizing inputs exploring new
paths, coverage-guided fuzzers like AFL~\cite{JQF} discover
deep-seated bugs and vulnerabilities. Bohme {\em et
  al.}~\cite{10.1145/3133956.3134020} introduced coverage-based
greybox fuzzing as a Markov Chain, presenting AFLFast as an extension
of AFL that significantly increases path coverage. 
GreyOne~\cite{244046} is a coverage-guided greybox fuzzer that
incorporates data flow analysis to prioritize paths that are more
likely to lead to vulnerabilities. Pasareanu and
Visser~\cite{10.1145/3364452.3364455} surveyed new trends in symbolic
execution relevant to fuzz testing. Cadar {\em et
  al.}~\cite{10.5555/1855741.1855756} introduced KLEE, a symbolic
virtual machine built on top of LLVM that uses symbolic execution to
systematically explore various paths.



Recent advancements in fuzzing techniques have shown promise in
overcoming the coverage plateau issue.
CODAMOSA~\cite{10.1109/ICSE48619.2023.00085} leverages the synergy
between search-based software testing (SBST) and LLMs to push beyond
the coverage plateau. By integrating SBST with LLMs, CODAMOSA explores
the input space via the embeddings of input values to generate more
diverse and sophisticated unit tests. Fuzz4All~\cite{xia2024fuzz4all}
can work across multiple programming languages, combined with its
uses of LLMs for input generation. Although Fuzz4All can theoretically generate test cases for
incomplete code, it still requires the execution of these test cases,
making it unsuitable for incomplete code in RQ1 and RQ3.



\section{Conclusion and Implications}
\label{sec:conclusion}

{\bf Novelty.} This paper introduces {\tool}, an LLM-based, multi-agent fuzz testing framework for (in)complete code without actual execution. We found that direct simulation of an iterative testing framework (one LLM for test generation and another for code coverage prediction), few-shot learning, Chain-of-Thought prompting, achieve only sub-optimal performance.
{\tool} introduces a novel iterative framework combining a {\bf multi-agent, two-phase feedback loop} with a {\bf condition-based dual-prompting strategy}. Those components are integrated into an iterative fuzzing loop for static runtime error detection—especially for incomplete code—is novel.


{\tool} generates significantly fewer test cases while detecting more runtime errors compared to the conventional fuzzer Jazzer~\cite{jazzer} and LLM-based Fuzz4All~\cite{xia2024fuzz4all}, with a precision of 58\% and a recall of 41\%. Furthermore, we demonstrate its effectiveness in detecting runtime errors in actual SO code snippets. {\tool} also supports for both Java and Python.


\vspace{2pt}
\noindent {\bf Impacts and Implications.} One could expand the capabilities of {\tool} in
key areas. \underline{First}, one could explore incorporating other
programming languages to enhance its applicability across diverse
software projects. The feedback loop mechanism can be generalized by integrating more sophisticated models to better identify and retain high-coverage test cases. 
\underline{Second,} LLM's multilingual capabilities make it feasible to adapt {\tool} to other languages and polyglot projects with multiple languages by updating (1) prompt templates to reflect language-specific idioms (e.g., pointer arithmetic in C), and (2) the PE module to handle coverage and language-specific exception prediction for low-level behaviors.
\underline{Third}, another area of interest is optimizing the computational efficiency of {\tool} further, making it more scalable. \underline{Fourth}, we will investigate the possibility of integrating {\tool} with other automated testing tools and frameworks to create a more holistic and seamless testing environment. This would enhance the detection of more complex runtime errors. Other types of LLM agents could be built for that purpose. 
\underline{Fifth}, we need a novel approach to model the execution coverage for different programming paradigms such as web programming where the coverage can be counted on the web pages. \underline{Finally}, for vulnerabilities, one could build on our work to support fuzzing to reveal different types of attacks without execution.



\section{Data Availability}

Our data and code are publicly available~\cite{cerberus-website}.



\bibliographystyle{ACM-Reference-Format}

\bibliography{references,references-planning,references-varex1}

\end{document}